\documentclass[12pt,english]{article}
\usepackage{helvet}
\usepackage[T1]{fontenc}
\usepackage{babel}
\usepackage{graphics}
\usepackage[latin1]{inputenc}
\usepackage{epsf,epsfig,psfrag}
\usepackage{amssymb}

\makeatletter

\date{\today}

\providecommand{\LyX}{L\kern-.1667em\lower.25em\hbox{Y}\kern-.125emX\@}

\makeatother

\begin{document}
  \title{CP violation and CKM predictions from Discrete Torsion}
  \author{S.A.Abel and A.W.Owen}
 \date{\textit{IPPP, Centre for Particle Theory, Durham
  University, DH1 3LE, Durham, U.K.}}
  
  \maketitle

\begin{abstract}
\noindent We present a supersymmetric D-brane model that has CP spontaneously broken by discrete torsion. The low energy physics is largely independent of the compactification scheme and the k\"{a}hler metric has `texture zeros' dictated by the the choice of discrete torsion. This motivates a simple ansatz for the k\"{a}hler metric which results in a CKM matrix given in terms of two free parameters, hence we predict a single mixing angle and the CKM phase. The CKM phase is predicted to be close to $\frac{\pi}{3}$.
\end{abstract}

\section{Introduction and motivation}
CP violation is a curious aspect of beyond the Standard Model physics.
 Baryogenesis (or rather its absence in the Standard Model) almost
 certainly indicates additional sources of CP violation beyond that
 which has been established by experiment to exist in the CKM matrix~\cite{sm1,sm2,sm3}. On the
 other hand the absence of electric dipole moments (EDMs) seems to
 indicate that the additional CP violation must have a very constrained
 form, possibly dictated by a direct connection with flavour structure.

 String theory may be giving us an important clue as to the real nature
 of CP violation; in string theory, what we call 4 dimensional CP is
 actually a gauge transformation plus Lorentz rotation of the 10 dimensional
 theory~\cite{cpgauge,choi}. Thus string theory is one of a finite class of theories in
 which CP is a discrete gauge symmetry which can only be spontaneously
 broken~\cite{choi}. Encouraged by this observation, a number of authors have attempted
 spontaneously to break CP in the effective supergravity approximation
 to various string models~\cite{sugra1,sugra2,sugra3,sugra4,sugra5,sugra6,sugra7,sugra8}. However, in these models it has proven very
 difficult to find a satisfactory suppression of EDMs and flavour changing
 processes. The problem seems to be that whatever fields break CP and
 generate flavour structure also contribute to supersymmetry breaking
 (specifically, their auxilliary fields aquire non-zero F-terms). Thus
 the supersymmetry breaking `knows about' the flavour and CP structure,
 giving rise to the usual supersymmetric flavour and CP problems. 

 We think that these problems can be avoided if one instead establishes
 the flavour and CP structure at the string theory level rather than
 the supergravity level. This is because string theory allows us to
 maintain supersymmetry which should in turn allow us to separate CP
 and flavour from supersymmetry breaking. In this paper we take a first
 step in this direction by constructing a supersymmetric MSSM-like
 model which has a full flavour structure and broken CP. In a later
 paper we shall examine the consequences for supersymmetry breaking,
 although we shall make some preliminary observations here.

 Our framework will be the `bottom-up' approach to string model building
 developed in ref~\cite{botup}. This approach allows one to construct a local configuration
 of D-branes that reproduces most of the phenomenological features of
 the MSSM without having to worry about the global properties of the
 compactification. The source of spontaneous CP violation will be discrete
 torsion, a choice of orbifold group action on the B-field background, analogous to Wilson lines for
 gauge fields~\cite{sharpe}. One nice feature of discrete torsion is that
 the Yukawa couplings contain a non-trivial complex phase, with the 
 breaking of 4 dimensional CP appearing as a phase in the CKM matrix.
 With a rather simple ansatz dictated by our choice of discrete torsion (similar to `texture zero' models in the
 MSSM) we find a CKM matrix defined by two free parameters. As a result, we predict a single mixing angle and the CKM phase, both of which lie within current experimental limits. Furthermore, quark mass ratios are determined and are seen to be reasonably close to experimental values. 

 It turns out
 that the up-quark Yukawa couplings are hermitian. The general
 idea of hermiticity has often been proposed as a way of suppressing
 EDMs in softly broken supersymmetric models~\cite{herm1,herm2,herm3,herm4,herm5} but it has
 been quite difficult to achieve in field theory. Here it arises as
 a natural consequence of the form of the Yukawa couplings with discrete
 torsion, indicating a possible resolution of the supersymmetric CP problem.

The paper is organized as follows. In the following section we introduce 
the model which is a generalization of the 
discrete torsion models presented in ref~\cite{botup}. Particular attention is paid to 
the effect of discrete torsion on the superpotential and it's role in breaking 4 dimensional CP. We then proceed to a more phenomenological discussion, including Yukawa couplings and the CKM matrix.

\section{A D-Brane model with Discrete Torsion}
First, a brief descriptive overview is given, and then a more detailed account that concentrates on the form of the superpotential. 

\subsection{A brief overview}
\label{overview}
The model we are interested in contains the following ingredients,

\begin{itemize}
\item A non-compact internal space $\mathbb{C}^{3}$, containing a $\mathbb{Z}_{3}\times \mathbb{Z}_{M}\times \mathbb{Z}_{M}$ orbifold singularity.
\item Coincident D3-Branes whose world-volume theory has the gauge group $SU(3)_{C}\times SU(2)_{L}\times U(1)_{Y}$.
\item Mutually perpendicular D7-Branes\footnote{We call a D7-brane whose worldvolume is $z_{i}=0$ a D$7_{i}$-brane.}  to cancel tadpoles, each with gauge group $U(1)\times U(2)$.
\end{itemize}

The D3-branes and D7-Branes are located on the $\mathbb{Z}_{3}\times \mathbb{Z}_{M}\times \mathbb{Z}_{M}$ singularity, this breaks supersymmetry from N=4 down to N=1, allowing for chirality. The orbifold group leads to model with three generations of Higgs' in addition to the standard quark-lepton generations. Furthermore the $\mathbb{Z}_{M}\times \mathbb{Z}_{M}$ factor allows for a choice of discrete torsion (this choice is labelled by an integer s). The discrete torsion then gives rise to a complex phase in the superpotential, these facts will be discussed in more detail in the rest of the paper.
   
\begin{figure}
\centering
\includegraphics*[0mm,0mm][84mm,84mm]{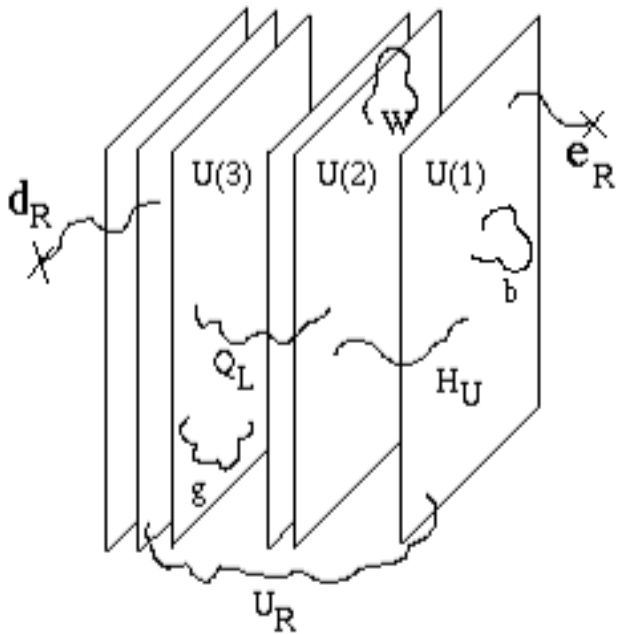}
\caption{The D3-branes are shown separated for clarity, although they are in fact coincident, and the D7-branes fill the figure. The position of end points of the strings determine how they transform under the SM gauge group, with fields such as the right-handed leptons and down quarks having one end on a D3 brane and another on a D7-brane.}
\label{branepic}
\end{figure}

Closed strings propagate only in the bulk and are gauge singlets, whereas the open strings are localized to the D-branes. The D3-branes are arranged in sets of 3s, 2s and s corresponding to the gauge groups U(3), U(2) and U(1)\footnote{The action of $\mathbb{Z}_{M}\times \mathbb{Z}_{M}$ on the string spectrum breaks the gauge group $U(sn)$ down to $U(n)$.}. Open strings with ends on D-branes within the same set give rise to the gauge bosons, and those with endpoints in different sets of D-branes give rise to matter fields. This makes it very simple to identify the MSSM fields as illustrated in figure~\ref{branepic}. The gauge group on the D3-branes is $U(3)\times U(2)\times U(1)$, remarkably however,  only one linear combination of the three U(1) factors is non-anomalous and this combination gives rise to the correct hypercharge assignments. This results in a world-volume theory with standard model gauge group.

\subsection{Orbifold action on the string spectrum}
\label{details}
A system of overlapping D3-branes and D7-branes has the following massless open string states in the spectrum,
\begin{enumerate}
\item In the `untwisted' 33 sector,
   \begin{itemize}
   \item world-volume gauge bosons $\lambda_{33}\psi^{\mu}_{-1/2}|0>$,
   \item complex scalars\footnote{These are the collective coordinates describing the embedding of the D3-branes in the internal space.} $\lambda_{33}\psi^{i}_{-1/2}|0>$, i=1,2,3.
   \item Weyl fermions $\lambda_{33}|s_{1},s_{2},s_{3},s_{4}>, s_{i}=\pm1/2$ and $\sum_{i=1}^{4}s_{i}$ odd. With $s_{4}$ determining the space-time chirality.
   \end{itemize}
\item In the `twisted' $37_{i}$ sector,
   \begin{itemize}
   \item complex scalars $\lambda_{37_{i}}|s_{j},s_{k}>$, $j,k\notin \{i,4\}$ and $s_{j}=s_{k}$,
   \item Weyl fermions $\lambda_{37_{i}}|s_{i},s_{4}>$, $s_{i}=s_{4}$.
   \end{itemize}
\end{enumerate}
Here $\lambda_{ij}$ are the Chan-Paton matrices. 

The action of the orbifold on these states can be split up into it's
action on the Chan-Paton matrices and the action on the string
worldsheet fields . For a $\mathbb{Z}_{N}$ orbifold with generator $\theta$ and
twist vector $v_{\theta}=(a_{1},a_{2},a_{3})/N$, N=1 supersymmetry
requires $a_{1}+a_{2}+a_{3}=0$. The generator $\theta$ acts on the 33
sector by leaving the gauge boson invariant, and transforming the
scalars and fermions respectively as,
\begin{eqnarray}
   \theta^{k}\psi^{i}_{-1/2}|0> &=& e^{2\pi i v_{\theta}^{i}k}\psi^{i}_{-1/2}|0>,\\ 
   \theta^{k}|\vec{s}> &=& e^{2\pi i \vec{s}.\vec{v_{\theta}}}|\vec{s}>.
\end{eqnarray}
Here $\vec{s}=(s_{1},s_{2},s_{3})$, and the twisted $37_{i}$ states
transform as the fermions above if we set any $s_{i}$ not defined to
be zero. For our particular case we take the twist vectors for $\mathbb{Z}_{3}$
and the two $\mathbb{Z}_{M}$ factors to be,
\begin{eqnarray}
v_{\theta} &=& (1,1,-2)/3 \nonumber,\\
v_{\omega_{1}}&=& (1,-1,0)/M \nonumber, \\
v_{\omega_{2}} &=& (0,1,-1)/M,
\end{eqnarray}
respectively.
The action of $\theta$ on the Chan-Paton matrices of the twisted and untwisted 
states respectively is,
\begin{eqnarray}
\theta\lambda_{33} &=& \gamma_{\theta,3}\lambda_{33}\gamma_{\theta,3}^{\dagger}, \\
\theta\lambda_{37_{i}} &=& \gamma_{\theta,3}\lambda_{37_{i}}\gamma_{\theta,7_{i}}^{\dagger},
\end{eqnarray}
where 
\begin{eqnarray}
\gamma_{\theta,3} &=& diag(I_{3s},\alpha I_{2s},\alpha^{2}I_{s}), \\
\gamma_{\theta,7_{i}}&=& diag(\alpha I_{s},\alpha^{2} I_{2s}).
\end{eqnarray}
Here we have defined $\alpha=e^{\frac{2\pi i}{3}}$. The $\gamma_{\theta}$ matrices form a unitary representation of
$\mathbb{Z}_{3}$ determined by our requirement of a SM gauge group 
and the tadpole cancelation condition,
\begin{equation}
\sum_{i=1}^{3}Tr(\gamma_{\theta,7_{i}})+3Tr(\gamma_{\theta,3})=0.
\end{equation}
The action of the generators $\omega_{i}$ is the same as
above, however this time the $\gamma_{\omega_{i}}$ matrices form a
projective representation of $\mathbb{Z}_{M} \times
\mathbb{Z}_{M}$~\cite{disctor,orientdis}, a possibility that arises since the action of
$\mathbb{Z}_{M}\times \mathbb{Z}_{M}$ allows for a choice of discrete torsion. This is defined in terms of a cocycle $\beta(g,h) \in H^{2}(\mathbb{Z}_{M}\times
\mathbb{Z}_{M},U(1))\cong \mathbb{Z}_{M}$~\cite{vafa,witten}, which manifests itself in the
projective representation where $\gamma_{g}\gamma_{h}=\beta(g,h)\gamma_{h}\gamma_{g}$. It follows that there are M non-equivalent choices of discrete torsion, 
described by distinct cocycles $\beta(\omega_{1},\omega_{2})=e^{2\pi i n/M}=\varepsilon$ where n=1,..,M. Such possibilities are conventionally distinguished
by an integer $s=M/gcd(n,M)$. For a string endpoint residing on a generic set of $sn^{(i)}$ D-branes, the action on the Chan-Paton factors is determined by the matrices,
\begin{eqnarray}
\gamma_{\omega_{1}}=\oplus_{l,m}\omega_{M}^{l}\hat{\gamma}_{\omega_{1}}\otimes I_{n^{(i)}_{lm}}, \\
\gamma_{\omega_{2}}=\oplus_{l,m}\omega_{M}^{m}\hat{\gamma}_{\omega_{2}}\otimes I_{n^{(i)}_{lm}},
\end{eqnarray}
where $\omega_{M}=e^{\frac{2 \pi i}{M}}$, $\sum_{l,m}n^{(i)}_{lm}=n^{(i)}$, $l,m=0,\ldots,(\frac{M}{s})-1$, and
\begin{eqnarray}
\hat{\gamma}_{\omega_{1}}=diag(1,\varepsilon^{-1},\ldots,\varepsilon^{-(s-1)}), \\
\hat{\gamma}_{\omega_{2}}= \left( \begin{array}{ccccc}
                        0 & 1 & 0 & \ldots & 0 \\
                        0 & 0 & 1 & \ldots & 0 \\
                  \vdots  & \ddots & \ddots & \ddots & \\
                        0 & 0 & 0 & \ldots & 1 \\
                        1 & 0 & 0 & \ldots & 0 
                    \end{array} \right)_{s \times s}
\end{eqnarray}
Since $Tr(\gamma_{\omega_{i}})=0$ tadpoles are unaffected by discrete torsion, it follows that there are no restrictions on the $n^{(i)}_{lm}$. For simplicity, we choose,
\begin{equation} 
n^{(i)}_{lm}=\left\{ \begin{array}{ll}
                     n^{(i)} & \mbox{if l=m=0}, \\
                      0    & \mbox{otherwise}.
                  \end{array}  \right.
\end{equation}

Combining the above transformations results in the projection equations 
which determine the spectrum and gauge group of the model to be that discussed 
above. For example, in the 33 sector we have,

\begin{equation}
\label{projeqns}
\begin{array}{lll}
\underline{\lambda_{33}\psi^{\mu}_{-1/2}|0>} & & \\ 
\lambda_{33}=\gamma_{\theta,3}\lambda_{33}\gamma_{\theta,3}^{\dagger} & \lambda_{33}=\gamma_{\omega_{1},3}\lambda_{33}\gamma_{\omega_{1},3}^{\dagger} & \lambda_{33}=\gamma_{\omega_{2},3}\lambda_{33}\gamma_{\omega_{2},3}^{\dagger} \nonumber \\
\underline{\lambda_{33}\psi^{1}_{-1/2}|0>} & & \\ 
\lambda_{33}=e^{2\pi i/3}\gamma_{\theta,3}\lambda_{33}\gamma_{\theta,3}^{\dagger} & \lambda_{33}=e^{\frac{2\pi i}{M}}\gamma_{\omega_{1},3}\lambda_{33}\gamma_{\omega_{1},3}^{\dagger} & \lambda_{33}=\gamma_{\omega_{2},3}\lambda_{33}\gamma_{\omega_{2},3}^{\dagger} \\
\underline{\lambda_{33}\psi^{2}_{-1/2}|0>} & & \\ 
\lambda_{33}=e^{2\pi i/3}\gamma_{\theta,3}\lambda_{33}\gamma_{\theta,3}^{\dagger} & \lambda_{33}=e^{-\frac{2\pi i}{M}}\gamma_{\omega_{1},3}\lambda_{33}\gamma_{\omega_{1},3}^{\dagger} & \lambda_{33}=e^{\frac{2\pi i}{M}}\gamma_{\omega_{2},3}\lambda_{33}\gamma_{\omega_{2},3}^{\dagger} \\
\underline{\lambda_{33}\psi^{3}_{-1/2}|0>} & & \\ 
\lambda_{33}=e^{-4\pi i/3}\gamma_{\theta,3}\lambda_{33}\gamma_{\theta,3}^{\dagger} & \lambda_{33}=\gamma_{\omega_{1},3}\lambda_{33}\gamma_{\omega_{1},3}^{\dagger} & \lambda_{33}=e^{-\frac{2\pi i}{M}}\gamma_{\omega_{2},3}\lambda_{33}\gamma_{\omega_{2},3}^{\dagger}. 
\end{array}
\end{equation}
\noindent Similar projection equations hold for superpartners and the $37_{i}$ sector. 

\subsection{The shift formalism and the spectrum}

A particularly useful formalism for finding the multiplets is the shift formalism 
of ref.\cite{orientdis,shift1,shift2,ofolds}
which makes use of the fact that the basis we have chosen for the gamma's is the Cartan basis 
(i.e. it is the basis where the diagonal elements correspond to the generators 
of the Cartan subalgebra). We write the gamma embeddings more explicitly\footnote{Previously 
we chose $n_{0}=3$, 
$n_{1}=2$ and $n_{2}=1$ to obtain the SM gauge group but here we leave them general.}
\begin{eqnarray}
 \gamma _{\theta,3 } & = & diag[(1)^{sn_{0}},(\alpha )^{sn_{1}},(\alpha
 ^{2})^{sn_{2}}]\\
 \gamma _{\omega _{1},3} & = & diag[(1,\varepsilon ^{-1},...\varepsilon
 ^{-(s-1)})^{n_{0}},(1,\varepsilon ^{-1},...\varepsilon
 ^{-(s-1)})^{n_{1}},(1,\varepsilon ^{-1},...\varepsilon ^{-(s-1)})^{n_{2}}]\\
 \gamma_{\omega _{2},3} & = & blockdiag[\hat{\gamma}_{\omega_{2}}I_{n_{0}},\hat{\gamma}_{\omega_{2}}I_{n_{1}},\hat{\gamma}_{\omega_{2}}I_{n_{2}}].
 \end{eqnarray}
 An obvious similar structure holds for the $\gamma_7$'s which we will not show
 here. Now the diagonal \( \gamma  \) matrices we write in terms of shift
 vectors that act on the root vectors of the states \( \rho ;
 \)
 \begin{eqnarray}
 \gamma _{\theta,3 } & = & e^{2\pi iV_{\theta }.H},\\
 \gamma _{\omega _{1},3} & = & e^{2\pi iV_{\omega _{1}}.H,}
 \end{eqnarray}
 whereas (for this choice of torsion) the effect of \( \gamma _{\omega _{2}} \)
 is a pure permutation of the root elements. 
 For the gauge states the projections are (mod 1) 
\begin{eqnarray}
\label{proj}
 \rho .V_{\theta } & = & 0\\
 \rho .V_{\omega _{1}} & = & 0\\
 \Pi _{\omega _{2}}(\rho ) & \rightarrow  & \rho ,
\end{eqnarray}
 where the final equation means that we must build a linear combination
 of the remaining states which is mapped into itself under permutations.
 The following linear combinations  of roots satisifes eq.(\ref{proj})
 \begin{eqnarray}
 \rho = &
 [\underline{((\pm  0..0)(\mp ..0)(0..0)^{n_{0}-2}})(0..0)^{n_{1}} (0..0)^{n_{2}}]
 & \\
 & +\, \, \, 
 [\underline{((0\pm 0..0)(0 \mp 0..0)(0..0)^{n_{0}-2}})(0..0)^{n_{1}}(0..0)^{n_{2}}] + ...
\end{eqnarray}
 \begin{eqnarray}
 \rho = &
 [\underline{(0..0)^{n_{0}}((\pm 0..0)(\mp ..0)(0..0)^{n_{1}-2}}) (0..0)^{n_{2}}]
 & \\
 & +\, \, \, 
 [\underline{(0..0)^{n_{0}}((0\pm 0..0)(0\mp 0..0)(0..0)^{n_{1}-2}})(0..0)^{n_{2}}] + ...
\end{eqnarray}
 \begin{eqnarray}
 \rho = &
 [\underline{(0..0)^{n_{0}} (0..0)^{n_{1}}((\pm 0..0)(\mp ..0)(0..0)^{n_{2}-2}})]
 & \\
 & +\, \, \, 
 [\underline{(0..0)^{n_{0}}(0..0)^{n_{1}}((0\pm 0..0)(0\mp 0..0)(0..0)^{n_{2}-2}})] + ...
\end{eqnarray}
 where the underlining means all permutations of brackets.
 Once we add the Cartan generators we find gauge fields in the adjoint
 representation of \( U(n_{0})\times U(n_{1})\times U(n_{2}) \). For
 the chiral states the projections are (mod 1) \begin{eqnarray*}
 \rho .V_{\theta } & = & v_{\theta,i}\\
 \rho .V_{{\omega_1}} & = & v_{\omega_1,i}\\
 \Pi _{\omega _{2}}(\rho ) & \rightarrow  & e^{2\pi iv_{\omega_2,i}}\rho .
 \end{eqnarray*}
 The multiplets are all in the bifundamental in this case and are
 \begin{equation}
 (\overline{n_{0}},n_{1},1)+(1,\overline{n_{1}},n_{2})+(n_{0},1,\overline{n_{2}})=
(\overline{3},2,1)+(1,\overline{2},1)+(3,1,\overline{1});
\end{equation}
i.e. with $n_{0}=3$, $n_{1}=2$ and $n_{2}=1$, we find a left-handed quark doublet, a right-handed quark singlet and a Higgs doublet. As an example consider bifundamental \( (\overline{n_{0}},n_{1},1) \)
 states coming from the 3$^{rd}$ complex plane, \( i=3 \). To satisfy the
 projection equations the roots must be the following linear combination 
 \begin{eqnarray}
\label{bifunds}
 \rho = &
 [\underline{(-0..0)(0..0)^{n_{0}-1}}\underline{(+0..0)(0..0)^{n_{1}-1}}(0..0)^{n_{2}}]
 & \\
 & +\, \, \, \varepsilon
 [\underline{(0-0..0)(0..0)^{n_{0}-1}}\underline{(0+0..0)(0..0)^{n_{1}-1}}(0..0)^{n_{2}}]
 & \\
 & +\, \, \, \varepsilon
 ^{2}[\underline{(00-0..0)(0..0)^{n_{0}-1}}\underline{(00+0..0)(0..0)^{n_{1}-1}}(0..0)^{n_{2}}]
 & +...
 \end{eqnarray}
Because the projection equations
for the $v_{\omega_2,i}$ are the same for all $i$, we find the same multiplets from each complex 
plane, and so for these models we have three generations.

\subsection{The superpotential}
The form of the superpotential can be deduced as in~\cite{D=4orbifold} and includes the following terms,
\begin{equation}
\label{superpotential} 
\label{super}
W=\varepsilon_{abc}Tr(\Phi^{a}_{3}\Phi^{b}_{3}\Phi^{c}_{3})+\sum_{i=0}^{2}Tr(\Phi^{i}_{3}\Phi^{37_{i}}\Phi^{37_{i}})+ \ldots, 
\end{equation} 
where $\Phi^{a}_{3}=\lambda_{33}^{a}\varphi^{a}_{3}$ and $\Phi^{37_{i}}=\lambda_{37_{i}}\varphi^{37_{i}}$
, with $\varphi^{a}_{3}$ a 33 chiral superfield and $\varphi^{37_{i}}$ a $37_{i}$ chiral superfield. Using the expressions derived for the Chan-Paton factors in appendix~\ref{appchan}, we have,
\begin{equation} \begin{array}{ll}
W & =\varepsilon_{abc}Tr((X_{33}^{(a)}\otimes Y^{(a)})(X_{33}^{(b)}\otimes Y^{(b)})(X_{33}^{(c)}\otimes Y^{(c)}))\varphi^{a}_{3}\varphi^{b}_{3}\varphi^{c}_{3}+... \\
  &=\varepsilon_{abc}Tr(X_{33}^{(a)}X_{33}^{(b)}X_{33}^{(c)})Tr(Z^{a}Z^{b}Z^{c})+ \ldots,
\end{array} \end{equation}
where $Z^{i}=Y^{(i)}\varphi^{i}_{3}$. Redefining the $Y^{(i)}$'s by a constant such that a factor of $Tr(X_{33}^{(1)}X_{33}^{(2)}X_{33}^{(3)})$ is absorbed into each term and noting that 
\begin{equation} \frac{Tr(X_{33}^{(2)}X_{33}^{(1)}X_{33}^{(3)})}{Tr(X_{33}^{(1)}X_{33}^{(2)}X_{33}^{(3)})}=e^{-2\pi i/M},\end{equation}
we arrive at,
\begin{equation}
\label{Upphase}
W=\epsilon_{abc}Tr(Z^{a}Z^{b}Z^{c})+(1-e^{-2\pi i/M})s_{abc}Tr(Z^{a}Z^{b}Z^{c})+...
\end{equation}
Here $s^{132}=s^{321}=s^{213}=1$ and all other components are zero. Identifying our supersymmetric SM fields, equations~(\ref{super}) and~(\ref{Upphase}) give us the yukawa type couplings,
\begin{equation}
W=h_{abc}\overline{u}^{a}_{R}Q^{b}_{L}H^{c}_{u}+f_{abc}\overline{d}^{a}_{R}Q^{b}_{L}H^{c}_{d}+\ldots,
\end{equation} 
where $h_{abc}=\epsilon_{abc}+(1-e^{-2\pi i/M})s_{abc}$ and $f_{abc}=\delta_{ab}\delta_{bc}$. 
Hence we have an up-quark Yukawa term containing a complex phase which gives rise to CP violation, and a prediction for the phase of the CKM matrix is 
presented in detail in section~\ref{CKM}. Before deriving the CKM matrix 
however, let us discuss the nature of the CP violation.

 \subsection{More on CP violation}

\begin{figure}
 
 {\centering
 \resizebox*{0.5\textwidth}{0.4\textheight}{\includegraphics{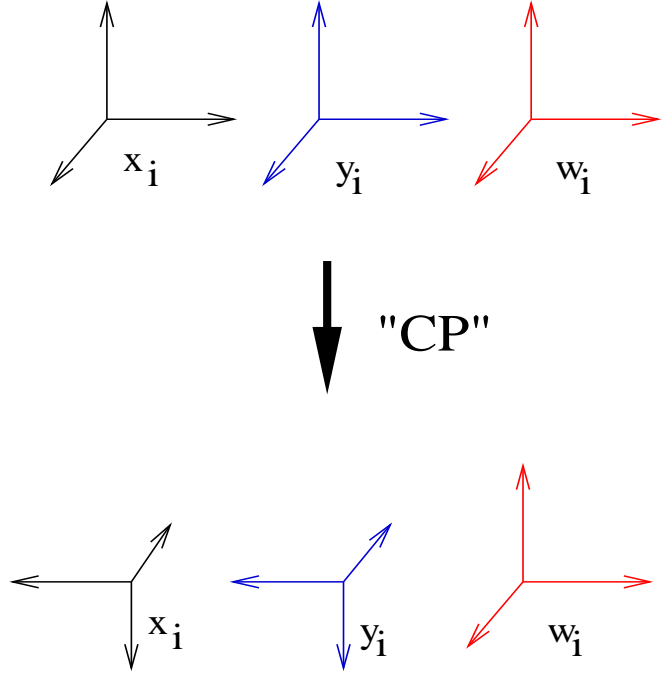}} \par}

 \caption{Rotations on 3 non-compact plus 3 compact directions corresponding
 to CP.\label{branes} }
 \end{figure}

 In the heterotic string, it is well known that a four dimensional
 CP conjugation corresponds to a rotation in the full 10 dimensional
 theory, and thus 4 dimensional CP turns out to be a discrete 
gauge symmetry~\cite{cpgauge,choi} (i.e.
 part of the 10D Poincare group). The situation is heuristically as
 shown in fig \ref{branes}. The three non-compact space dimensions
 are labelled \( x_{i} \), with \( y_{i} \) and \( w_{i} \) labelling
 6 internal compact dimensions. Parity is defined
 as a reflection in one direction, \( x_{1} \) say. By a rotation
 through \( \pi  \) in the \( x_{2}x_{3} \) plane, four dimensional parity is equivalent
 to a reflection in all the space directions, the more conventional
 definition (valid when the number of space dimensions is odd). However
 if we simultaneously reflect in an odd number of internal compact
 dimensions the total is a rotation in three orthogonal planes. 
 Complexifying the internal space
 as \begin{equation}
 z_{i}=w_{i}+iy_{i},\end{equation}
 we can, without loss of generality, choose the additional reflection
 corresponding to a CP transformation to act on the \( y_{i} \) coordinates,
 with \begin{equation}
 {\rm CP}
\, :\, x^\mu \rightarrow x_\mu,\, z_{i} \rightarrow z_{i}^{*}.\end{equation}
 By supersymmetry we must also reflect the fermionic superpartners,
 so that the transformation reverses the R-charges of the spectrum.
 In turn, the requirement of modular invariance is only satisfied if
 we reverse the gauge charges as well. (In the standard embedding this
 is a simple copying of the internal reflections on the space-time
 side into the \( SU(3) \) subgroup on the gauge side.) The nett result
 of the combined rotation plus gauge transformation is a 4 dimensional
 CP transformation.  

 An even simpler picture holds for branes at orbifold fixed points, as we
 shall now see, and it is the geometric configuration of torsion plus
 singularity that breaks CP. 
 As an example, consider the chiral multiplets written explicitly 
in eq.(\ref{bifunds}). The corresponding antiparticles
 are given by the conjugate excitation, and the projections are reversed
 accordingly;
\begin{eqnarray}
 \overline{\rho }.V_{\theta } & = & -v_{\theta,i}\\
 \overline{\rho }.V_{\omega _{1}} & = & -v_{\omega_1,i}\\
 \Pi _{\omega _{2}}(\overline{\rho }) & \rightarrow  & e^{-2\pi
 iv_{\omega_2,i}}\overline{\rho },
 \end{eqnarray}
 giving \begin{eqnarray}
\label{anticomb}
 \overline{\rho }= &
 [\underline{(+0..0)(0..0)^{n_{0}-1}}\underline{(-0..0)(0..0)^{n_{1}-1}}(0..0)^{n_{2}}]
 & \\
 & +\, \, \, \varepsilon
 ^{-1}[\underline{(0+0..0)(0..0)^{n_{0}-1}}\underline{(0-0..0)(0..0)^{n_{1}-1}}(0..0)^{n_{2}}]
 & \\
 & +\, \, \, \varepsilon
 ^{-2}[\underline{(00+0..0)(0..0)^{n_{0}-1}}\underline{(00-0..0)(0..0)^{n_{1}-1}}(0..0)^{n_{2}}]
 & +...
 \end{eqnarray}
 as expected. 

 Now consider the rotation \( x^{\mu }\rightarrow x_{\mu }, \) \
  \( z_{i}\rightarrow z_{i}^{*}. \)
 Again by supersymmetry we transform \( \psi _{i}\rightarrow \overline{\psi
 _{i}} \)
 and need to reverse the chiralities in the Ramond sector. 
The particle projections now trivially become antiparticle
 projections. To get the right antiparticles however, we also need to ensure
 that we have the same projective representation which is the 
case if the discrete torsion is unchanged (e.g. so that we have 
 {\em same} linear combination as in eq.(\ref{anticomb}). 
If this were not the 
case we would simply be unable to identify these particular 
rotations with CP conjugation.) For open string sectors the 
 effect of the torsion in a sector with $\mathbb{Z}_M\times \mathbb{Z}_M$ twists given by 
($g\equiv a,b$ ; $h\equiv a',b'$) can be written as 
\begin{equation}
\beta(g,h)=\epsilon^{ab'-ba'}.
\end{equation}
 The effect of the \( z_{i}\rightarrow z_{i}^{*} \) reflection on the 
torsion in any sector is simply to reverse the twists,
 \( (a,b\, ;\, a',b'\, ) \) \( \rightarrow -(a,b\, ;\, a',b'\, ) \)
 so $\beta(g,h)$ is indeed unchanged and the rotation gives us the 
antiparticle. i.e. 4 dimensional CP conjugation does indeed correspond
 to a rotation in the 10D theory, with or without discrete torsion. 

 It is precisely because the discrete 
 torsion is invariant under CP conjugation that the superpotential
 interactions are not. Indeed the superpotential is of the form
\begin{equation}
 W=Tr(\Phi _{1}\Phi _{2}\Phi _{3}+\varepsilon^{-1/n} \Phi _{2}\Phi _{1}\Phi _{3}),\end{equation}
where here the $\Phi$'s stand for generic chiral fields, and where a summation 
over the SU(3) Levi-Cevita symbol is implied. 
 Under the CP transformation we conjugate the chiral superfields but the 
discrete torsion remains the same. Thus \begin{equation}
 \int d\theta ^{2}\, W+\int d\overline{\theta }^{2}\, \overline{W}\, \rightarrow
 \, \int d\overline{\theta }^{2}\, \overline{W}_{CP} + \int d\theta ^{2}W_{CP} \,
 \end{equation}
 where\begin{equation}
 W_{CP}=Tr(\Phi _{1}\Phi _{2}\Phi _{3}+\varepsilon ^{1/n}\Phi _{2}\Phi _{1}\Phi
 _{3}).\end{equation}
 So it is the discrete torsion term in the superpotential that is directly 
responsible for breaking 4 dimensional CP and, because CP 
conjugation is associated with a 10D Lorentz rotation, it is a
spontaneously broken symmetry. We will now divert our attention to the more phenomenological issues.

\section{Yukawas and the CKM matrix}
\label{CKM}
In this section we first discuss some theoretical considerations for
obtaining the canonical yukawa couplings. In particular, we derive bottom-up
properties of the k\"{a}hler metric that have not been previously noted.
We then proceed to the CKM matrix and predictions for the CKM angles and
quark mass ratios. Finally we discuss radiative 
corrections between the string and weak scales.

\subsection{Normalisation of Yukawas}
\subsubsection{Bottom-up properties of the k\"{a}hler metric}
To be consistent with N=1 supersymmetry, we assume the low-energy effective field theory decribing our visible sector is an N=1 SUGRA theory.
It is well known that in such a theory scalar fields have a non-canonical kinetic energy term given by a k\"{a}hler metric which is a function of the compactification moduli. This normalistion of scalar fields, 
results in an effective normalisation of coupling constants. In particular, if  we expand the k\"{a}hler potential to first order in the charged
matter fields\footnote{i.e. fields charged under our visible sector gauge group.}, we have
\begin{equation}
\mbox{K}=\mbox{\^{K}}(M^{a},M^{a*})+\mbox{\~{K}}(M^{a},M^{a*})_{\alpha \beta}Tr(\Phi^{\alpha *}\Phi^{\beta})+...,
\end{equation}
where $M^{a},M^{a*}$ are the closed string moduli fields and we obtain,
\begin{equation}
\label{noreqn}
\hat{h}_{abc}=e^{\hat{K}/2}h_{lmn}(\mbox{\~{K}}^{-1/2})^{l}{}_{a}(\mbox{\~{K}}^{-1/2})^{m}{}_{b}(\mbox{\~{K}}^{-1/2})^{n}{}_{c},
\end{equation}
where $\hat{h}_{abc}$ are the canonically normalised yukawa couplings. It is important to note that $\mbox{\~{K}}_{\alpha \beta}$ factorises into different components, one for each sector of the spectrum, e.g. in our case the 33 and 37 sector fields have the seperate k\"{a}hler matter metrics, $(\mbox{\~{K}}_{33})_{\alpha\beta}$ and $(\mbox{\~{K}}_{37})_{\alpha \beta}$.

One of the strengths of the bottom-up approach is that, up until now, our discussion has been independent of any compactification scheme. In particular, the gauge group, SM spectrum and 3 generations of matter fields are all features independent of the global structure of the model. This affords us two approaches to the determination of the yukawas,
\begin{itemize}
  \item Select a particular compactification and calculate the exact k\"{a}hler potential.
  \item Choose an ansatz for $\mbox{\~{K}}_{33}$ and $\mbox{\~{K}}_{37}$ (note that the CKM matrix is independent of $\mbox{\^{K}}$).
\end{itemize}
The first approach is technically difficult and only acheivable in the
simplest of cases, e.g. toroidal orbifolds. In addition, it goes against the
bottom-up approach, which is to constrain the model as much as possible from
the local rather than global properties. The second
approach, at first sight, seems too arbitrary. However, the choice of
possible ans\"{a}tze is restricted by two observations. Firstly, the discrete
torsion can constrain the form of $\mbox{\~{K}}_{33}$. Such a situation
arises as a consequence of $Tr(\Phi_{33}^{\alpha *}\Phi_{33}^{\beta})$
containing a factor $Tr((X_{33}^{(\alpha)})^{\dagger}X_{33}^{(\beta)})$
arising from the Chan-Paton factors. This matrix contains `texture zeros', dependent on the discrete torsion(n), which are inherited by $(\mbox{\~{K}}_{33})_{\alpha \beta}$. Note that the k\"{a}hler metric is hermitian, and we will coventionally refer only to the zeros in the upper triangular section of $\mbox{\~{K}}_{33}$. There are three possible cases,
\begin{itemize}
\item minimal discrete torsion, i.e. $n=1$, with $\mbox{\~{K}}_{33}$ diagonal,
\item $n=2$, with two off diagonal zeros,
\item and $n \geq 3$, with no off diagonal zeros.
\end{itemize}
These results are derived in the appendices.

Finally, we have the restriction that $\mbox{\~{K}}_{37}$ is diagonal. $(\mbox{\~{K}}_{37})_{ij}$ can be calculated from tree-level modulus-matter scattering~\cite{kaplu}. The amplitude is given by a disc diagram with two open string vertex operators on the boundary, a $(37_{i})$ and $(7_{j}3)$ vertex, and two closed string vertex operators attached to the interior, corresponding to the moduli fields. However vertices of $(37_{i})$ and $(7_{i}3)$ strings must come in pairs in order to get a consistent D-brane boundary on the disc. It follows that in this case i=j and hence $\mbox{\~{K}}_{37}$ is diagonal.

This allows us to explore possibilities for interesting flavour structure and CP violation in our yukawas without worrying about the exact compactification scheme we are using. That is, we choose an ansatz for the k\"{a}hler metric based on our choice of discrete torsion, not our choice of compactification. 

\subsubsection{Explicit expressions for Yukawas}
Defining the hermitian matrices,
\begin{equation}
\begin{array}{l}
t_{33}=\mbox{\~{K}}_{33}^{-1}, \\ 
t_{37}=\mbox{\~{K}}_{37}^{-1}, 
\end{array}
\end{equation}
and substituting into~(\ref{noreqn}) we obtain, in the general case, canonical up-quark yukawas,
\begin{equation}
\begin{array}{l}
\hat{h}^{abc}=e^{\hat{K}/2}h_{lmn}(t_{33}^{1/2})^{la}(t_{33}^{1/2})^{mb}(t_{33}^{1/2})^{nc} \\
=e^{\hat{K}/2}(\varepsilon_{lmn}+(1-e^{-2\pi i/M})s_{lmn})(t_{33}^{1/2})^{la}(t_{33}^{1/2})^{mb}(t_{33}^{1/2})^{nc}.
\end{array}
\end{equation}
Furthermore giving the three generation of Higgs fields vevs, $\langle h_{u}\rangle_{c}$, we obtain
\begin{equation}
(Y_{u})^{ab}=e^{\hat{K}/2}(\varepsilon_{lmn}+(1-e^{-2\pi i/M})s_{lmn})(t_{33}^{1/2})^{la}(t_{33}^{1/2})^{mb}(t_{33}^{1/2})^{nc}\langle h_{u}\rangle_{c}.
\end{equation}
This expression also holds if we associate the actual physical Higgs fields with a linear combination of the $H_{u/d}^{a}$
fields. That is
\begin{equation} 
\hat{H}_{u/d}^{i}=U_{u/d}^{ij}\tilde{H}_{u/d}^{j}, 
\end{equation}
where $\hat{H}_{u/d}^{i}$ are the physical Higgs fields and $U_{u/d}$
is a unitary matrix. Then assuming that only $\hat{H}^{1}_{u/d}$ is
light and the other two physical Higgs' are heavy, we simply replace $\langle h_{u}\rangle_{c}$ with $(U_{u}^{\dagger})_{c1}$.

Similarly, for the down-quark yukawas, we obtain,
\begin{equation}
(Y_{d})^{ab}=e^{\hat{K}/2}\sum_{l=1}^{3}(t_{33}^{1/2})^{la}(t_{37}^{1/2})^{lb}(t_{37}^{1/2})^{lc}\langle h_{d}\rangle_{c}.
\end{equation}
Using these formulae we can calculate the CKM matrix and determine the possiblities for CP violation in our model.

\subsection{The CKM matrix}
\subsubsection{A simple ansatz from a choice of discrete torsion}
Using the bottom-up properties of our k\"{a}hler metric, that is,
\begin{itemize}
\item the zeros in $\mbox{\~{K}}_{33}$ are determined by the discrete torsion,
\item $\mbox{\~{K}}_{37}$ is diagonal,
\end{itemize}
we try the following simple ansatz, take M=5, n=2  and
\begin{equation} t_{33}^{\frac{1}{2}}= \left( \begin{array}{ccc}
                                               \eta\alpha\varepsilon^2 & 0 & \alpha \varepsilon \\
                                                  0    & 1 & 0 \\
                                              \alpha \varepsilon    &   0    & \alpha
                                          \end{array} \right), \end{equation}
\begin{equation} 
t_{37}^{\frac{1}{2}}= diag[1,\varepsilon,1], 
\end{equation}
\begin{equation} \begin{array}{ll}
       \langle h_{u}\rangle= & (\varepsilon,1,0), \\
       \langle h_{d}\rangle= & (1,\varepsilon,\varepsilon).
   \end{array} \end{equation}
Notice that our choice of discrete torsion introduces two off diagonal zeros in $\mbox{\~{K}}_{33}$ which are preserved in $t_{33}^{\frac{1}{2}}$.

Using this ansatz we can calculate the Yukawas and CKM matrix. The Yukawas have the following form,
\begin{equation}
\label{upyuk}
Y_{u}=\alpha^{2} \varepsilon \left( \begin{array}{lcl}
                               (1-e^{-2i \pi/5})\eta \varepsilon^{2} & (1-e^{-2i \pi/5})\eta \varepsilon^{3} & (1-\eta e^{-2i \pi/5})\varepsilon \\
                                         (1-e^{-2i \pi/5})\eta \varepsilon^{3} & 0 &  (\eta-e^{-2i \pi/5})\varepsilon^{2} \\
                                         (\eta-e^{-2i \pi/5})\varepsilon &  (1-\eta e^{-2i \pi/5})\varepsilon^{2} &  (1-e^{-2i \pi/5})
                             \end{array} \right),
\end{equation}
and
\begin{equation}
\label{downyuk}
Y_{d}=\alpha \varepsilon \left( \begin{array}{lll}
                                 \eta \varepsilon & 0 & \varepsilon \\
                                         0   & \frac{\varepsilon^2}{\alpha} & 0 \\
                                        1   &   0 & 1
                              \end{array} \right).
\end{equation}
With the freedom to rephase the right handed quark fields (because the model
is supersymmetric we can make rephasings just as in the Standard Model),
$Y_{u}$ can be put into hermitian form. Notice that $\alpha$ appears only as
an overall factor in $Y_{u}$ and in a single entry in $Y_{d}$, as a
consequence the CKM matrix is only dependent on the two parameters
$\varepsilon$ and $\eta$, with $\alpha$ determining the mass of the down
quark. Up to overall factors, which include the prefactors of~(\ref{upyuk}) and~(\ref{downyuk}), the masses of the quarks are given approximately for small $\eta$ and
$\varepsilon$ by,

\begin{equation}
\begin{array}{lll}
m_{d}=\frac{\varepsilon^{2}}{\alpha}, &
m_{s}=\frac{\varepsilon(1-\eta)}{\sqrt{2}}, & m_{b}=
\sqrt{2}+\frac{1}{4\sqrt{2}}\varepsilon^{2}+\frac{1}{20\sqrt{2}}\eta\varepsilon^{2},\\
m_{u}=1.9\eta\varepsilon^{4}, &
m_{c}=\sqrt{\frac{2}{3}}\varepsilon^{2}-\frac{\sqrt{26}}{3}\eta\varepsilon^{2} +\sqrt{\frac{2}{3}}\eta^{2}\varepsilon^{2},&
m_{t}=\frac{2}{\sqrt{3}}+\sqrt{\frac{2}{3}}\varepsilon^{2}-\frac{1}{\sqrt{3}}\eta\varepsilon^{2}.
\end{array}
\end{equation}

The CKM matrix can be written approximately in terms of these mass eigenvalues, but it is not very illuminating to show the functional dependence here. Instead, in the next subsection, we depict graphically it's functional dependence on $\varepsilon$ and $\eta$.

\subsubsection{Predictions: CKM angles and Quark mass ratios} 
In the standard parametrisation, the CKM matrix is given in terms of three
 mixing angles, $\sin{\theta_{12}}$, $\sin{\theta_{23}}$,
 $\sin{\theta_{13}}$, and a complex phase $\delta$. Figures 3 to 6 illustrate
 the dependence of these CKM angles on $\varepsilon$ and $\eta$. The blue
 shaded areas highlight the regions where the CKM angles agree with
 experiment~\cite{datagroup}. Since we are discussing string scale
 predictions we must also include RGE effects in running from the string to
 the weak scale. This effect is discussed in the next subsection, where we
 estimate that $\sin{\theta_{23}}$ and $\sin{\theta_{13}}$ are suppressed by
 approximately $30\%$. This is depicted in the figures by a red shaded region
 which higlights values of $\varepsilon$ and $\eta$ that will reproduce
 experimental values after taking into account RGE effects. This information
 is summarised in figure~\ref{agreegrph} where the regions in parameter space
 which produce a realistic CKM matrix are highlighted. In particular, it can
 be seen that $\delta$ is approximately $\pi/3$.

The quark mass ratios are also determined, $\frac{m_{d}}{m_{b}}$ can be fixed
by choosing a value for $\alpha$, while the rest of the quark mass ratios are
dependent only on $\varepsilon$ and $\eta$. Average values for the blue and
red regions are presented in table~\ref{masses} and compared to experimental
values. As can be seen, we have good agreement with experiment except for
$\frac{m_{u}}{m_{c}}$ which is too large. Note that we have not included RGE
effects here, which would tend to increase the value of
$\frac{m_{c}}{m_{t}}$. As we have chosen such a constrained and simple
ansatz, we consider the close agreement of the CKM angles and quark mass
ratios with experiment to be successful. Furthermore, it has not been
necessary to introduce any large artificial hierarchies in our ansatz
(which would require similar hierarchies in the corresponding moduli fields), to achieve mass eigenvalues and mixing angles close to those of the Standard Model.
\begin{figure}[!hp]
\centering
\includegraphics*[10mm,0mm][160mm,84mm]{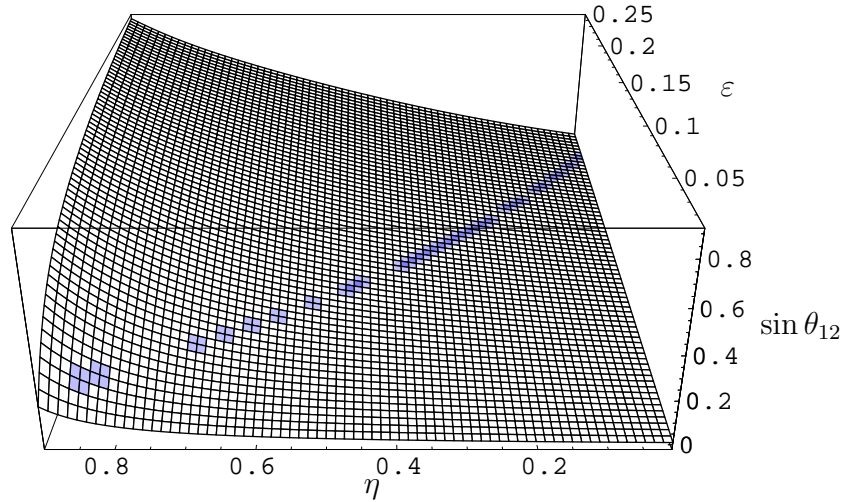}
\put(-80,60){$\sin{\theta_{12}}$}
\put(-230,0){$\eta$}
\put(-95,150){$\varepsilon$}
\caption{$\sin{\theta_{12}}$ as a function of $\varepsilon$ and $\eta$.}
\label{s12graph}
\end{figure}

\begin{figure}[!hp]
\centering
\includegraphics*[10mm,0mm][160mm,84mm]{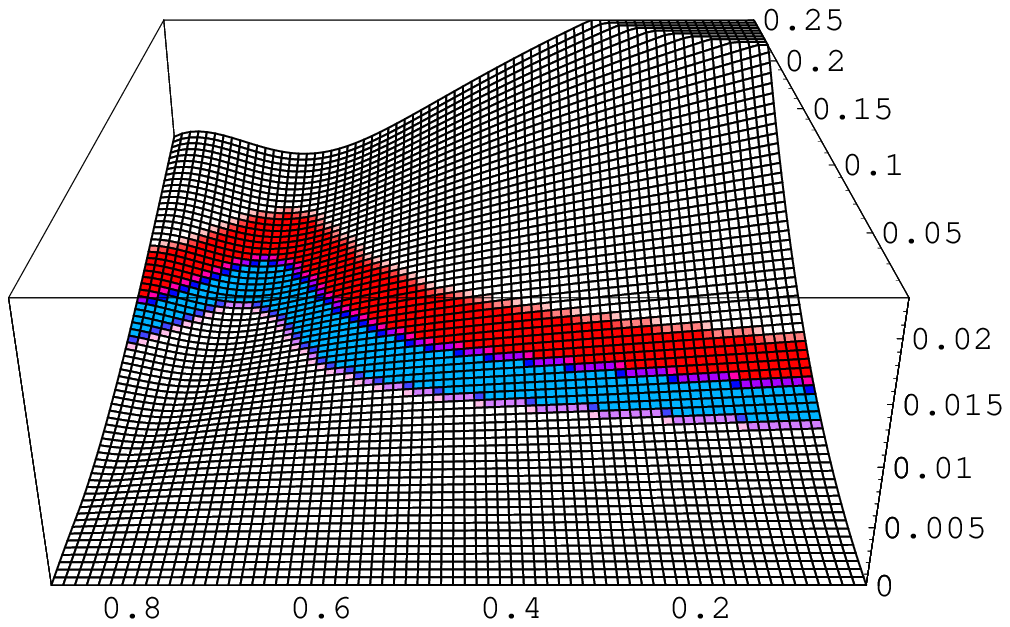}
\put(-80,60){$\sin{\theta_{13}}$}
\put(-230,0){$\eta$}
\put(-100,150){$\varepsilon$}
\caption{$\sin{\theta_{13}}$ as a function of $\varepsilon$ and $\eta$.}
\label{s13graph}
\end{figure}

\begin{figure}[!hp]
\centering
\includegraphics*[10mm,0mm][160mm,84mm]{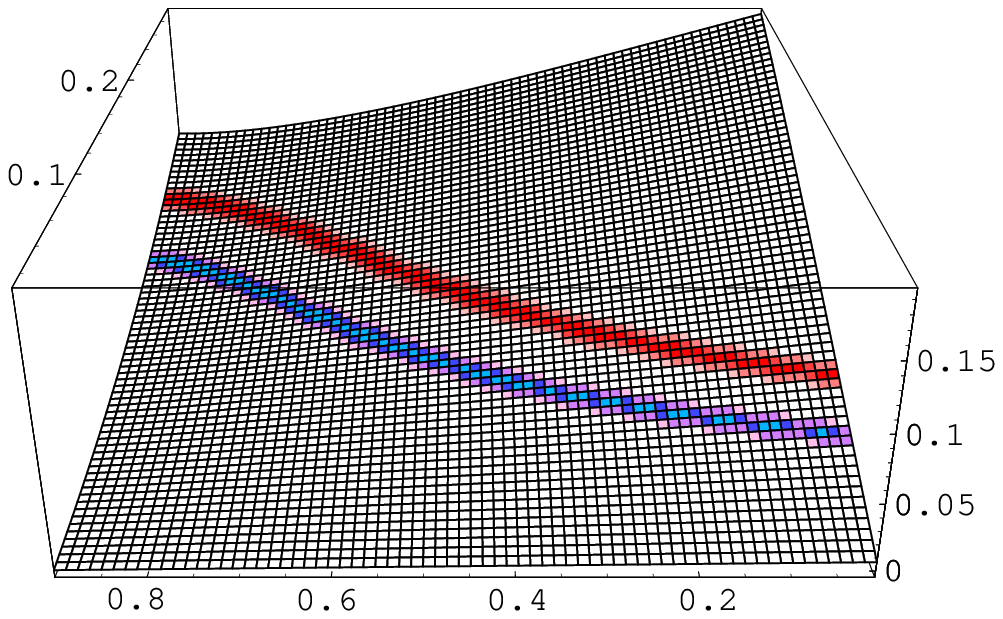}
\put(-80,60){$\sin{\theta_{23}}$}
\put(-220,0){$\eta$}
\put(-370,150){$\varepsilon$}
\caption{$\sin{\theta_{23}}$ as a function of $\varepsilon$ and $\eta$.}
\label{s23graph}
\end{figure}

\begin{figure}[!hp]
\centering
\includegraphics*[10mm,5mm][160mm,100mm]{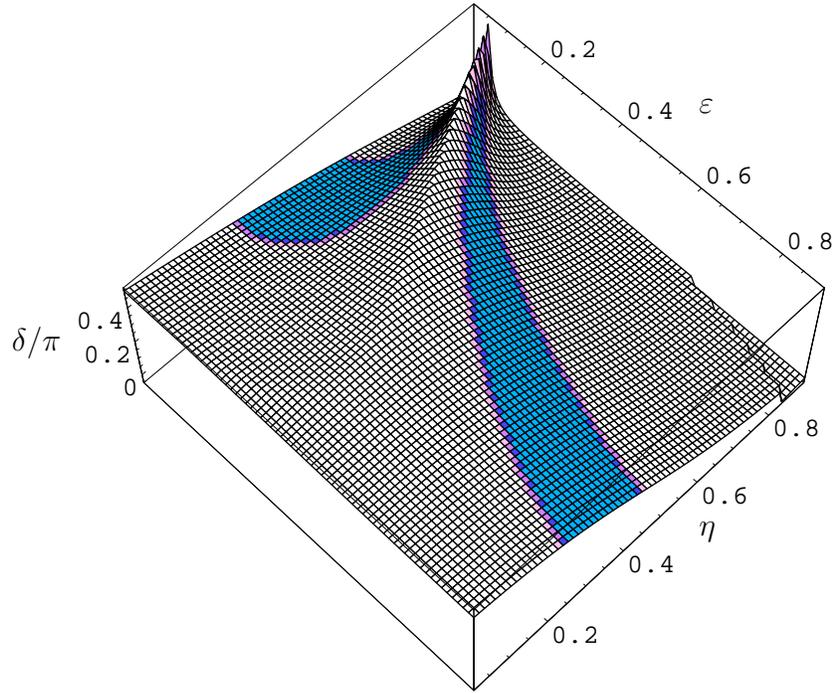}
\put(-390,130){$\delta/\pi$}
\put(-130,220){$\varepsilon$}
\put(-130,60){$\eta$}
\caption{$\delta/\pi$ as a function of $\varepsilon$ and $\eta$.}
\label{delgraph}
\end{figure}

\begin{figure}[!hp]
\centering
\includegraphics*[10mm,5mm][160mm,100mm]{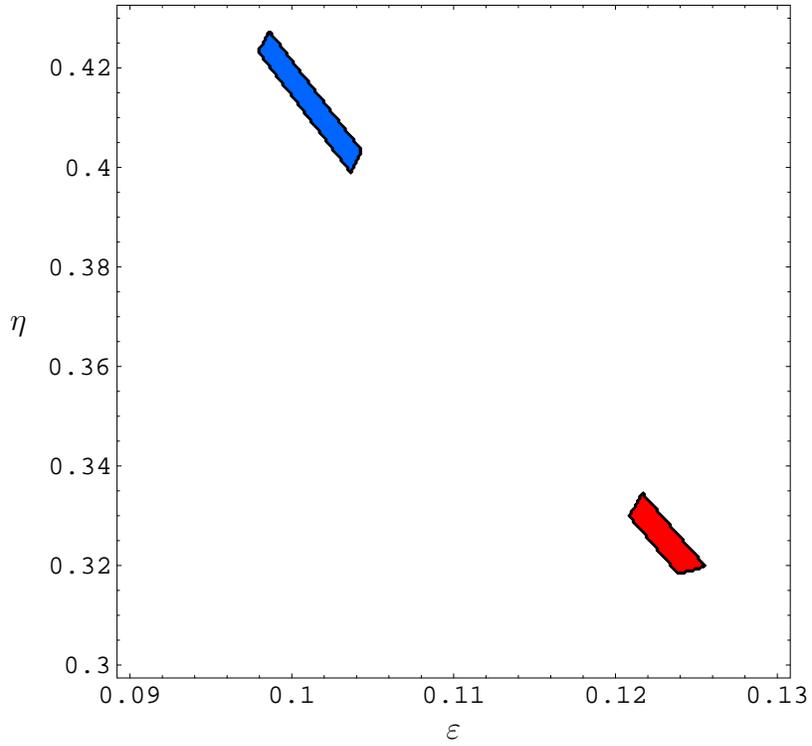}
\put(-380,145){$\eta$}
\put(-215,-10){$\varepsilon$}
\caption{The shaded regions highlight the values of $\varepsilon$ and $\eta$ which give an agreement with experimental data. The blue/red region without/with taking into account RGE effects.}
\label{agreegrph}
\end{figure}
  
\begin{table}[!hp]
\centering
 {\centering \begin{tabular}{|c|c|c|c|}
 \hline 
 Mass Ratio&
 Blue Region&
 Red Region&
 Experimental Value\\
 \hline
 \hline 
 \( \frac{m_{c}}{m_{t}} \)&
 $2.5 \times 10^{-3}$ &
 $5 \times 10^{-3}$ &
 $5.57-8.27 \times 10^{-3}$\\
 \hline
 \( \frac{m_{s}}{m_{b}} \)&
 $0.029$ &
 $0.041$&
 $0.018-0.0387$\\
 \hline
 \( \frac{m_{u}}{m_{c}} \)&
 $2.4\times 10^{-2}$&
 $2.1\times 10^{-2}$&
 $1.07-4.5 \times 10^{-3}$\\
 \hline
 \end{tabular}\par}
\caption{Quark mass ratios}
\label{masses} 
\end{table}

\pagebreak

\subsubsection{RGE effects}
\label{rge}

 The CKM matrix is expected to be effected by renormalization group
 running. For instance, when we are close to the so-called quasifixed
 point, the top Yukawa is large at the GUT or string scale
 and runs to lower fixed point values, and hence RGE effects are important. 

 \begin{figure}[h]
   \centering
   \psfrag{Exact}[bl][bl]{$\;\;\;\textrm{\large{Angle}}$}
   \psfrag{Approx}[bl][bl]{$\textrm{\large{Angle}}_{\textrm{\tiny{GUT}}}$}
   \epsfig{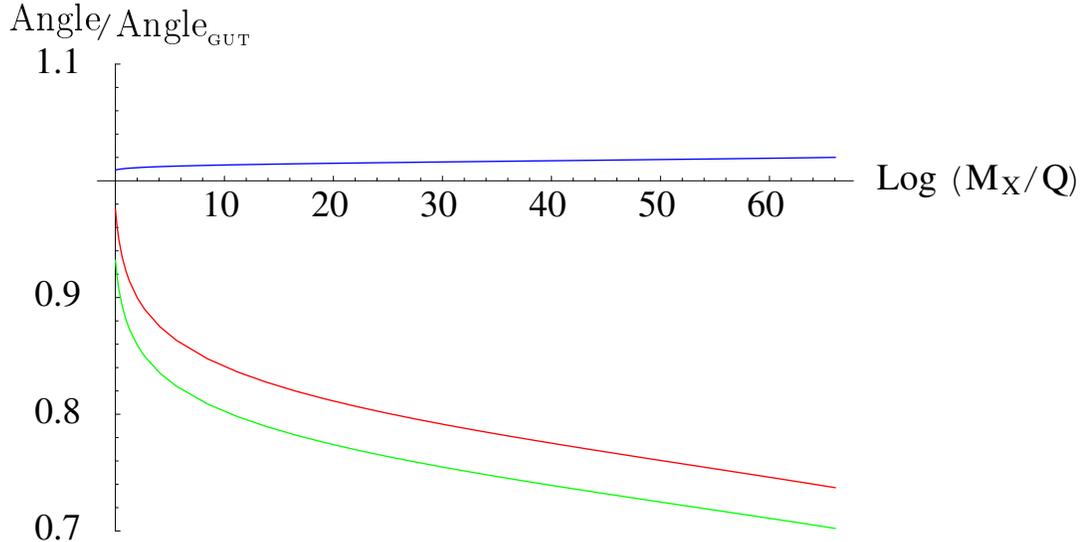}
   \caption{\label{ckmrun} CKM running from \( M_{X}\approx 2\times
     10^{16}GeV \) to \( M_{W}\) for a top Yukawa of $\lambda_t=5$ at
     the GUT scale.  Where red
     $\equiv\frac{\sin\theta_{23}}{(\sin\theta_{23})_{GUT}}$, blue
     $\equiv\frac{\frac{\sin\theta_{13}}{\sin\theta_{12}}}{(\frac{\sin\theta_{13}}{\sin\theta_{12}})_{GUT}}$,
     green $\equiv\frac{\delta}{\delta_{GUT}}$. $\sin\theta_{12}$
     does not run significantly.}
 \end{figure}
 
However, the effect on the CKM angles is generally small, as can be seen by
 inspection of the RG equation. Figure~\ref{ckmrun} shows how the numerical values of
 the CKM parameters change with renormalization scale when the top quark 
 Yukawa is close to the quasifixed point. The CKM angles are 
 suppressed by the running and from the figure we see that 
 the maximum reasonable suppression of the $\theta_{23}$ and 
 $\theta_{13}$ angles is $\approx 30\% $. The phase $\delta$ does not change 
 significantly and is expected to be close to $\frac{\pi}{3}$.

\section{Summary and discussion}
 We have presented a bottom-up supersymmetric D-brane model with phenomenologically 
 viable CP violation, broken by discrete torsion.
 Furthermore, a simple assumption about the form of the k\"{a}hler metric, motivated by a choice of discrete torsion, produces a CKM matrix described by
 only two free parameters. As a consequence, we predict a single CKM mixing angle and the CKM phase to be close to $\pi/3$, both of which lie within current experimental limits. We believe this to be the first such prediction from any string model.

 We believe that this class of models may be a first step towards a 
 solution of the SUSY flavour and CP problems. Generally, these 
 problems arise because supersymmetry breaking ``knows'' about CP and 
 flavour. The approach that we favour is therefore to use 
 the `bottom-up' approach to build a supersymmetric MSSM, complete 
 with all flavour and CP structure, {\em before} discussing 
 supersymmetry breaking. 

 An interesting hint may lie in the fact 
 that the Yukawa structure in these models 
 seems to favour a hermitian flavour structure. Hermiticity has 
 been suggested as a way to combat large EDMs and solve the SUSY 
 CP problem~\cite{stringcp}. In our analysis only the up-quark Yukawas are hermitian and so we cannot yet claim to have a solution. However we think that the appearance of 
 hermitian flavour structure is intriguing.

 It would be interesting to investigate the dependence of these results
 on our choice of discrete torsion (i.e. our choice of $n$), and the resulting ansatz for the k\"{a}hler metrics.
 Finally an analysis of the possibilities for a phenomenologically viable breaking of supersymmetry has been well motivated. All of these 
 questions will be the subject of future work.

\section{Acknowledgements}
We would like to thank Shaaban Khalil for assistance with computations. S.A.Abel is supported by a PPARC Opportunity grant and A.W.Owen by a PPARC studentship.

\appendix
\begin{center}
\textbf{\LARGE{Appendix}}
\end{center}
\section{Calculation of Chan-Paton factors}
\label{appchan}
Consider a generic chiral superfield whose Chan-Paton factor, as a result of the $\gamma_{\theta}$ projection, is of size $sn^{(i)}\times sn^{(j)}$. If we act on this state with an element $g=(a,b)=\omega_{1}^{a}\omega_{2}^{b}\in \mathbb{Z}_{M}\times \mathbb{Z}_{M}$, we require~\footnote{This is just a more general version of the projection equations.}
\begin{equation}
\label{A}
\lambda_{sn^{(i)}\times sn^{(j)}}^{(k)}=r^{(k)}(a,b)\gamma_{g,3}\lambda_{sn^{(i)}\times sn^{(j)}}^{(k)}\gamma_{g,3}^{\dagger},
\end{equation}
where $r^{(k)}(a,b)$ is a phase from the action on the string worldsheet fields and for a string endpoint on a set of $sn^{(i)}$ D-branes,
\begin{equation}
\label{form}
\gamma_{g}=\oplus_{l,m}(\sqrt{\epsilon})^{-ab}(\omega_{M}^{l}\gamma_{\omega_{1}})^{a}(\omega_{M}^{m}\gamma_{\omega_{2}})^{b}\otimes I_{n^{(i)}_{lm}},
\end{equation}
where $\omega_{M}=e^{2 \pi i/M}$, $\sum_{l,m} n^{(i)}_{lm}=n^{(i)}$ and $l,m=0,...,(M/s)-1$. It follows that we can factor the Chan-Paton matrix by writing,
\begin{equation}
\label{B}
\lambda^{(k)}_{sn^{(i)}\times sn^{(j)}}=\oplus_{l,m}X^{(k)}_{s\times s}\otimes Y^{(k)}_{n^{(i)}_{lm}\times n^{(j)}_{lm}}.
\end{equation}
where $X^{(k)}$ is determined by the action of $\mathbb{Z}_{M}\times \mathbb{Z}_{M}$ and $Y^{(k)}$ is left undetermined. Substituting~(\ref{B}) into~(\ref{A}) removes the dependence on $Y^{(k)}$ giving,
\begin{equation}
\label{C}
X^{(k)}\gamma_{\omega_{1}}^{a}\gamma_{\omega_{2}}^{b}=r^{(k)}(a,b)\gamma_{\omega_{1}}^{a}\gamma_{\omega_{2}}^{b}X^{(k)}.
\end{equation}
Since the $\gamma_{\omega_{i}}$ matrices form a projective representation and hence satisfy $\gamma_{g}\gamma_{h}=\beta(g,h)\gamma_{h}\gamma_{g}$, we can solve~(\ref{C}) with the ansatz $X^{(k)}=\gamma_{\omega_{1}}^{p_{k}}\gamma_{\omega_{2}}^{q_{k}}$. This results in,
\begin{equation}
\label{simple}
\beta(\omega_{1},\omega_{2})^{p_{k}b-q_{k}a}=r^{(k)}(a,b).
\end{equation}
For 33 sector fields we have,
\begin{equation}
r_{33}^{(k)}(a,b)=e^{2\pi i(av_{\omega_{1}}+bv_{\omega_{2}})},
\end{equation}
and for 37 sector fields we have,
\begin{equation}
r_{37}^{(k)}(a,b)=e^{\pi i (a\sum_{i \neq k}v_{\omega_{1}}^{i}+b\sum_{i\neq k}v_{\omega_{2}}^{i})}. 
\end{equation}
Substituting these expressions into~(\ref{simple}) we find,
\begin{equation}
\begin{array}{l}

p_{k}= \left\{ \begin{array}{ll}
                 \frac{a_{2}^{k}}{n} & \mbox{For 33 fields} \\
                 \frac{1}{2n}\sum_{i \neq k} a_{2}^{i} & \mbox{For 37 fields}
              \end{array} \right. , \\
q_{k}= \left\{ \begin{array}{ll}
                 -\frac{a_{1}^{k}}{n} & \mbox{For 33 fields} \\
                 -\frac{1}{2n}\sum_{i \neq k} a_{1}^{i} & \mbox{For 37 fields}
              \end{array} \right. ,\\
\end{array}
\end{equation}
where $v_{\omega_{i}}=\frac{1}{M}(a_{i}^{1},a_{i}^{2},a_{i}^{3})$. Finally we have, for the 33 fields,
\begin{equation}
\label{chans}
\begin{array}{lll}
X^{(1)}_{33}=\gamma_{\omega_{2}}^{-\frac{1}{n}}, & X^{(2)}_{33}=\gamma_{\omega_{1}}^{\frac{1}{n}}\gamma_{\omega_{2}}^{\frac{1}{n}}, & X^{(3)}_{33}=\gamma_{\omega_{1}}^{-\frac{1}{n}},
\end{array}
\end{equation}
and for the 37 fields,
\begin{equation}
\begin{array}{lll}
X_{37_{1}}=\gamma_{\omega_{2}}^{\frac{1}{2n}}, &  X_{37_{2}}=\gamma_{\omega_{1}}^{-\frac{1}{2n}}\gamma_{\omega_{2}}^{-\frac{1}{2n}}, & X_{37_{3}}=\gamma_{\omega_{1}}^{\frac{1}{2n}}. 
\end{array}
\end{equation}

\section{Determination of zeros in $\mbox{\~{K}}_{33}$}
\label{zeros}
As discussed in the main text, the zeros in $\mbox{\~{K}}_{33}$ are inherited from the hermitian matrix,
\begin{equation}
Z_{ij}=Tr((X^{(i)}_{33})^{\dagger}X^{(j)}_{33}).
\end{equation}
However, equation~(\ref{chans}) does not determine the $X^{(i)}_{33}$ uniquely. Therefore, we choose the $\mbox{n}^{th}$ principal root of
$\gamma_{\omega_{k}}$ defined by,
\begin{equation}
\gamma_{\omega_{k}}^{\frac{1}{n}}= S^{\dagger}(\gamma_{\omega_{k}}^{diag})^{\frac{1}{n}}S,
\end{equation}
with S the unitary matrix which diagonalises $\gamma_{\omega_{k}}$ and,
\begin{equation}
(\gamma_{\omega_{k}}^{diag})_{ij}=\delta_{ij}\lambda_{i}^{\frac{1}{n}}.
\end{equation}
Here $\lambda_{i}$ is the $i^{th}$ eigenvalue of $\gamma_{\omega_{k}}$, $\lambda_{i}^{1/n} \equiv e^{\frac{1}{n}\log(\lambda_{i})}$ and
we restrict $\arg(\lambda_{i})$ to the interval $(-\pi, \pi]$, thus taking the principal value of $\log(\lambda_{i})$.

With these choices, the $X^{(i)}_{33}$ are uniquely determined and we have,

\begin{equation}
\begin{array}{ll}
\label{z12}
Z_{12}= & \frac{1}{s}(\sum_{l=1}^{g(s)}e^{-4\pi i (l-1)/ns}+e^{4\pi i/n}\sum_{l=g(s)+1}^{s}e^{-4\pi i (l-1)/ns}) \\
        & \times(\sum_{p=0}^{p_{max}} e^{2 \pi i p/n}\sum_{u(p)> j \geq l(p)}e^{-2\pi i(j-1)/M}),
\end{array}
\end{equation}
\begin{equation}
\begin{array}{ll}
\label{z13}
Z_{13}= & \frac{1}{s}(\sum_{l=1}^{g(s)}e^{-2\pi i (l-1)/ns}+e^{2\pi i/n}\sum_{l=g(s)+1}^{s}e^{-2\pi i (l-1)/ns}) \\
        & \times(\sum_{p=0}^{p_{max}} e^{-2 \pi i p/n}\sum_{u(p)> j \geq l(p)}e^{2\pi i(j-1)/M}),
\end{array}
\end{equation}
and,
\begin{equation}
\begin{array}{ll}
\label{z23}
Z_{23}= & \frac{1}{s}(\sum_{l=1}^{g(s)}e^{2\pi i (l-1)/ns}+e^{-2\pi i/n}\sum_{l=g(s)+1}^{s}e^{2\pi i (l-1)/ns}) \\
        & \times(\sum_{p=0}^{p_{max}} e^{-4 \pi i p/n}\sum_{u(p) > j \geq l(p)}e^{4\pi i(j-1)/M}),
\end{array}
\end{equation}
where,
\begin{equation}
\begin{array}{l}
p_{max}=[\frac{(s-1)n}{M} + \frac{1}{2}], \\
u(p)=(2p+1)\frac{M}{2n}+1, \\
l(p)=(2p-1)\frac{M}{2n}+1, \\
g(s)= \left\{ \begin{array}{ll}
                k & \mbox{if s=2k,} \\
                k+1 & \mbox{if s=2k+1,}
               \end{array} \right.
\end{array}
\end{equation}
with $[..]$ denoting rounding to the nearest integer. Furthermore, since the  $X^{(i)}_{33}$ are unitary $s \times s$ matrices, we have
\begin{equation}
Z= \left( \begin{array}{ccc}
             s & Z_{12} & Z_{13} \\
             Z_{12}^{*} & s & Z_{23} \\
              Z_{13}^{*} & Z_{23}^{*} & s
          \end{array} \right).
\end{equation}
Finally, it can be easily seen from~(\ref{z12}),~(\ref{z13}) and~(\ref{z23})
that $Z_{12}=Z_{13}=Z_{23}=0$ for $n=1$, $Z_{12}=Z_{23}=0$ and $Z_{13} \neq 0$
for $n=2$ and for $n \geq 3$ all elements of Z are non-zero. Hence giving
rise to the three distinct cases for the distribution of zeros in
$\mbox{\~{K}}_{33}$ as described in the main text.

\bibliography{CPbib}
\bibliographystyle{unsrt}
\end{document}